\documentclass[preprint,aps,floatfix,superscriptaddress]{revtex4}

\usepackage{dcolumn}
 \usepackage{booktabs} 
 \usepackage{ulem}

\usepackage{graphicx}
\usepackage{amsfonts}
\usepackage{amsmath,bm,amssymb}
\usepackage[usenames]{color}

\def\vxc{V^{\rm xc}}
\def\ei{\varepsilon_i}

\def\ej{\varepsilon_j}

\begin{document}

\title{Quasiparticle self-consistent $GW$ study of cuprates:
  electronic structure, model parameters, and the two-band theory for
  T$_c$}

\author{Seung Woo Jang}
\affiliation{Department of Physics, Korea Advanced Institute of
  Science and Technology (KAIST), Daejeon 305-701, Korea }

\author{Takao Kotani} \affiliation{Department of Applied Mathematics
  and Physics, Tottori University, Tottori 680-8552, Japan}

\author{Hiori Kino} \affiliation{National Institute for Materials
  Science, Sengen 1-2-1, Tsukuba, Ibaraki 305-0047, Japan.}

\author{Kazuhiko Kuroki} 
\affiliation{Department of Physics, Osaka University,
  Machikaneyama-Cho, Toyonaka, Osaka 560-0043, Japan}

\author{Myung Joon Han} \email{mj.han@kaist.ac.kr}
\affiliation{Department of Physics, Korea Advanced Institute of
  Science and Technology (KAIST), Daejeon 305-701, Korea }
\affiliation{ KAIST Institute for the NanoCentury, Korea Advanced
  Institute of Science and Technology, Daejeon 305-701, Korea }

\begin{abstract}
Despite decades of progress, an understanding of unconventional
superconductivity still remains elusive. An important open question is about
the material dependence of the superconducting properties.
Using the quasiparticle self-consistent $GW$ method, we re-examine the
electronic structure of copper oxide high-T$_c$ 
materials. We show that QSGW captures several important features,
distinctive from the conventional LDA results. The energy level
splitting between $d_{x^2-y^2}$ and $d_{3z^2-r^2}$ is significantly
enlarged and the van Hove singularity point is lowered. 
The calculated results compare better 
than LDA with recent experimental results from resonant inelastic xray
scattering and angle resolved photoemission experiments.
This agreement with the experiments supports  
the previously suggested two-band theory for the
material dependence of the superconducting transition temperature, T$_c$.
\end{abstract}


\maketitle

%\section{Introduction}
After the seminal work of finding high temperature
superconductivity in a ceramic copper oxide material \cite{first1986},
efforts to understand the cuprate have long been a central part of modern
condensed matter physics.  Although many intriguing aspects of its
electronic behaviors have been unveiled and great progress has been
made, an understanding of its superconducting
mechanism, the novel interplay between competing phases, and 
its relationship to other correlated phenomena is still far from clear 
\cite{review9,review10,review11,review12,review13,
  review14,review15,review16,review17,review18,review19}. One
simple, well-defined but still open question is what determines the
superconducting transition temperature (T$_c$), or its material
dependency. For example, the T$_c$ of the single layer cuprates can be
different by a factor of two; $\sim$40K for La$_2$CuO$_4$ and $\sim$90K
for HgBa$_2$CuO$_4$. On the one hand, it may be too early to
ask this question, while the superconducting mechanism itself still
remains elusive. On the other, however, figuring out the detailed
features behind the material dependency can provide the crucial hint
for further understanding of superconductivity and other related
properties. 
In fact, many directly or indirectly related theoretical 
studies have been performed on this issue 
\cite{Kotliar,Weber,Millis,Hozoi,Honerkamp,Mori,
Ohta,Maekawa,Takimoto,Andersen_1995,Feiner,Pavarini,Hozoi2}.

In this regard, a notable suggestion has recently been made 
\cite{Sakakibara_PRL_2010,Sakakibara_PRB_2012_1,Sakakibara_PRB_2012_2,
  Sakakibara_PRB_2014}. According to this theory, the T$_c$ of the
single-layer cuprate can be described with a two-orbital 
model that considers both the $d_{x^2-y^2}$ and $d_{3z^2-r^2}$ Wannier orbitals,
and the energy level offset between the two orbitals,
$\Delta$E, plays a
key role in determining T$_c$.  Whereas the larger value of this
energy separation produces the higher T$_c$ ({\it e.g.,} the case of
HgBa$_2$CuO$_4$) due to the better one band feature achieved, the
smaller value results in the lower T$_c$ ({\it e.g.,} the case of
La$_2$CuO$_4$) in spite of the better nested Fermi surface.  The
calculated Eliashberg parameter ($\lambda$) based on the many-body
calculation using the fluctuating exchange (FLEX) approximation \cite{Bickers,Dahm} clearly
exhibits a linear dependence on $\Delta$E while the other parameters
are shown to be less important.  Further, this theory can be extended
to the bilayer case \cite{Sakakibara_PRB_2014},
which explains the 
correlation between the Fermi surface shape and T$_c$ \cite{Pavarini}.

%KK-start
%We note that the key parameters used in this type of theoretical
%approaches largely rely on the non-interacting band structure
%typically obtained by local density approximation (LDA). A widely used
%standard technique is to resort to LDA or its simple variant such as
%generalized gradient approximation (GGA) for the band structure
%calculation and then to extract the model parameters for many-body
%calculation via the downfolding based on maximally localized Wannier
%orbitals. This is adopted in the previous study mentioned above
%\cite{Sakakibara_PRL_2010,Sakakibara_PRB_2012_1,Sakakibara_PRB_2012_2,
%  Sakakibara_PRB_2014}.  Although it is one of the well-defined
%state-of-the-art techniques, it should also be pointed out that the
%description of non-interacting band structure by LDA or GGA has a
%clear limitation.  

A possible experimental test to verify this two-band theory is to examine the correlation 
between T$_c$ and $\Delta E$, the Fermi surface shape, or 
the partial density of states of the $d_{3z^2-r^2}$ orbital, which can be 
measured by recent techniques such as resonant inelastic
xray scattering (RIXS), and (angle resolved) photoemission spectroscopy (ARPES).
However, while the theoretical $\Delta E$ or the Fermi surface shape 
was obtained from the LDA and
used as the ``inputs'' for the many-body FLEX calculation 
in Refs.~\onlinecite{Sakakibara_PRL_2010,Sakakibara_PRB_2012_1,Sakakibara_PRB_2012_2,  Sakakibara_PRB_2014},  
 the experimentally determined
$\Delta E$ and the Fermi surface shape should be regarded as the ``outputs'' or ``results''
after the consideration of the many-body correlation effects beyond LDA/GGA.
In fact, while a RIXS study  
reports that T$_c$ is  higher for larger 
$\Delta E$ \cite{Sala},
 the actual experimental value of $\Delta E$ 
is larger than the theoretical evaluation, presumably due to this 
``input vs. output'' problem. 
One possible way to  resolve this problem, at  least partially, is to re-evaluate $\Delta E$ 
as an output of the FLEX calculation. 
However, this approach would suffer from various ambiguities regarding 
the Hubbard interaction strength and 
the definition of the renormalized $\Delta E$. It is problematic since 
a quantitative comparison is required in between the theory and experiment,
while only the qualitative comparison was made regarding
$T_c$ in Refs. \onlinecite{Sakakibara_PRL_2010,Sakakibara_PRB_2012_1,Sakakibara_PRB_2012_2,  Sakakibara_PRB_2014}.  

In the present paper, we use a first-principles approach, 
exploiting the quasiparticle self-consistent $GW$ (QSGW) method. 
 It enables us to take into 
account the correlation effects beyond LDA/GGA. In this way, we can 
obtain a well-defined renormalized $\Delta E$ without 
introducing adjustable parameters.

In Ref.~\onlinecite{Hozoi}, a quantum chemical approach was adopted to evaluate 
the energy level offset between Cu-$d_{x^2-y^2}$ and $d_{3z^2-r^2}$ orbitals, 
where the correlation effects were taken into account within 
a cluster-based configuration-interaction-type calculation. 
A good agreement with the RIXS experiment
was found by assuming the energy difference of ferromagnetically and 
antiferromagnetically ordered states to be $2J$, where $J$ is the 
antiferromagnetic coupling constant. 
Our approach is fairly different and is along the line of 
the first-principles band calculation as in Refs.~\onlinecite{Sakakibara_PRL_2010,Sakakibara_PRB_2012_1,Sakakibara_PRB_2012_2,  Sakakibara_PRB_2014}.
In the sense mentioned above, 
the calculated $\Delta E$ can be compared to 
the experiments, while it should not be regarded as an input parameter for 
the many-body calculation, because doing so 
would result in a partial 
double counting of the correlation effects. Still, the present approach 
can also provide a first-step hint toward obtaining a better ``non-interacting''
Hamiltonian that can be used as an input for the many-body calculation 
of superconductivity. In fact,  it is  known that the non-interacting 
Hamiltonian obtained from LDA
has a problem when used as an input for the FLEX calculation, and the  
LDA/GGA estimation of $\Delta E$  for La$_2$CuO$_4$ is 
too small to account for the maximum $T_c$ of 40K  in the 
La$_2$CuO$_4$ \cite{MiyaharaArita}.
In this context, it is worth pointing out that the 
$GW$ method has been successfully applied to
the many of strongly correlated materials in combination with, for
example, dynamical mean field theory (DMFT)
\cite{GW-DMFT-1,GW-DMFT-2,GW-DMFT-3,GW-DMFT-4}.

%						& \multicolumn{3}{c}{ E$_{e_g}$ (4Ds + 5Dt) (eV)} & \multicolumn{3}{c}{3Ds $-$ 5Dt (eV)} & \multicolumn{3}{c}{10Dq (eV)}& T$_{c}^{max}$ (K)\\	
%						& LDA & QSGW & Exp.\cite{Sala} & LDA & QSGW & Exp. \cite{Sala} & LDA & QSGW & Exp.\cite{Sala} &Exp.\\\hline
%	Sr$_2$CuO$_2$Cl$_2$	& 0.94 & 2.05 & 1.97 & 0.22 & 0.21 & 0.33 & 0.81 & 1.82 & 1.50 & 30 \cite{Liu}\\	
%	La$_2$CuO$_4$		& 0.84 & 1.66 & 1.70 & 0.40 & 0.11 & 0.32 & 1.47 & 2.13 & 1.80 & 40 \cite{Sakakibara_PRL_2010}\\ 
%	HgBa$_2$CuO$_4$		& 1.67 & 2.25 & None & 0.42 & 0.42 & None & 0.90 & 1.57 & None & 90 \cite{Sakakibara_PRL_2010}\\
%	CaCuO$_2$			& 2.04 & 2.46 & 2.65 & 0.01 & 0.43 & 0.31 & 0.95 & 1.54 & 1.64 & 110 \cite{Azuma}\\

\section*{Results and Discussion}

To our
knowledge, there is no previous QSGW study for the cuprate band structure
although it has been discussed conceptually \cite{Das-Bansil}.
Here we first examine the
electronic structure and the two-band theory for the material-dependent
T$_c$ of a single layer cuprate. While the QSGW
calculation produces notable differences in the band structure and
Fermi surface from LDA, the two-band explanation for T$_c$ still
remains valid.  QSGW results of model
parameters are presented and compared to the 
RIXS 
 data as well as the LDA calculations. It
clearly shows that the parameters produced by QSGW are in better agreement with
the experiment.
Finally, we investigate the epitaxially strained
La$_2$CuO$_4$ whose noticeable T$_c$ increase
has been previously reported. Two-band theory also works well
for this situation.

\subsection*{Electronic structure and the T$_c$ of single layer compounds}
\begin{figure*}[h]
\begin{center}
\begin{tabular}{c c}
\includegraphics[width=15cm,angle=0]{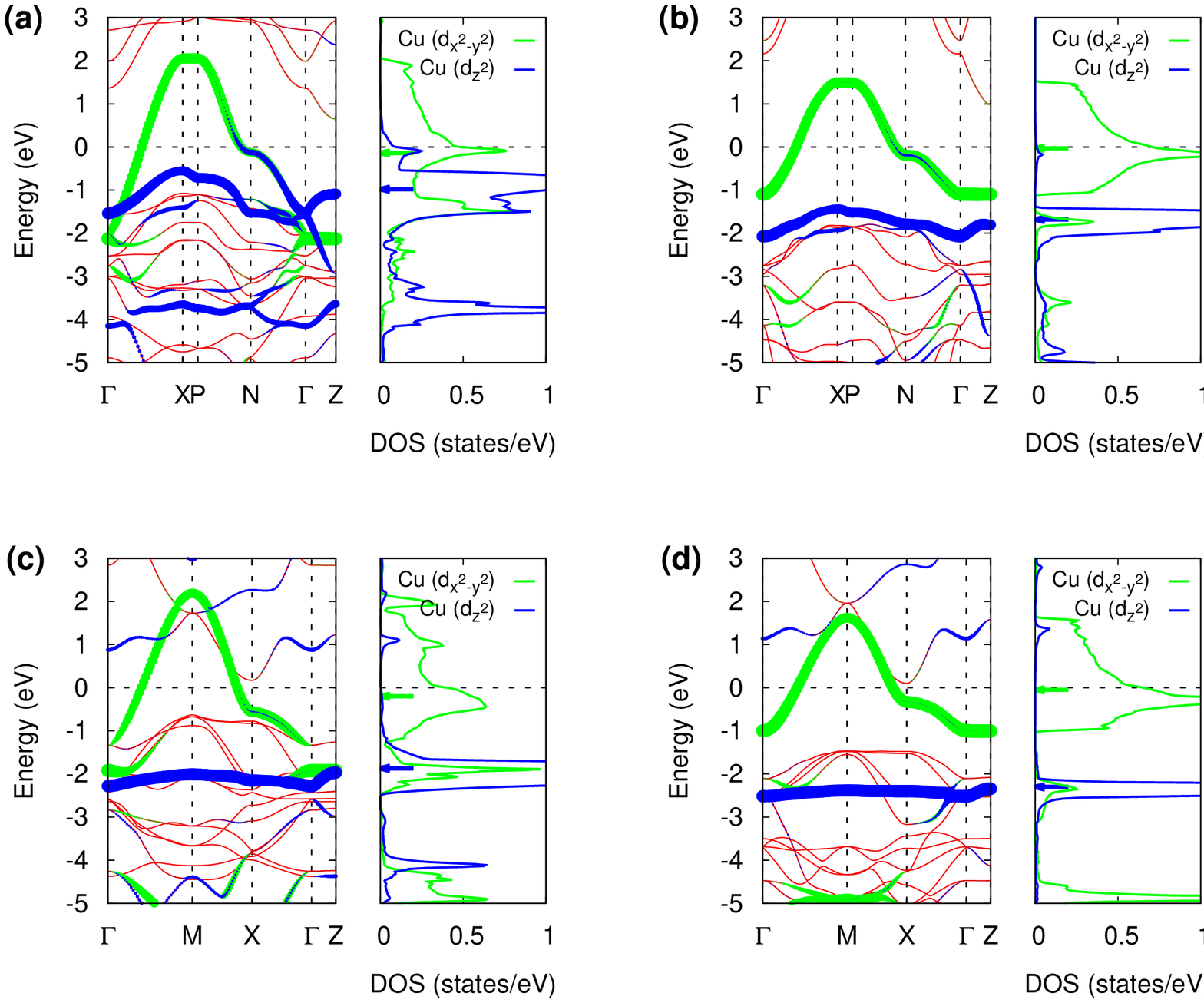}
\end{tabular}
\caption{% The band dispersion and PDOS of La$_2$CuO$_4$ (a, b)
%  and HgBa$_2$CuO$_4$ (c, d) calculated by (a, c) LDA and (b, d) QSGW.
%The green and blue colors refer
%  to the $d_{x^2-y^2}$ and $d_{3z^2-r^2}$ characters while the size of
%  the colored dots represents their weight. The same color scheme was
%  used for PDOS. Other bands than the two $e_g$ states are represented
%  by red color. The center of mass position of PDOS is marked by an
%  arrow. Fermi energy is set to be 0.
\label{Figure 1}}
%La214
\end{center}
\end{figure*}

Fig.~\ref{Figure 1}(a) and (b) show the band dispersion and projected density of states (PDOS) of
La$_2$CuO$_4$ calculated by LDA and QSGW, respectively.  The LDA
result is in good agreement with the previous calculation (see, for
example, Ref.~\onlinecite{Sakakibara_PRL_2010} and
Ref.~\onlinecite{Sakakibara_PRB_2012_1}). In the QSGW, several
important differences are noted. First, the band width of both $e_g$
orbitals are significantly reduced, by about 1.30 and 0.65 eV for
$d_{x^2-y^2}$ and $d_{3z^2-r^2}$, respectively, indicating that the
band width overestimation (or effective mass underestimation) problem
of LDA is somehow overcome by the QSGW procedure.  Another key difference
is that 
the separation between the $d_{x^2-y^2}$ and $d_{3z^2-r^2}$ bands
becomes larger in QSGW, as seen in Fig.~\ref{Figure 1}(b).
The $d_{x^2-y^2}$ energy level, E$_{x^2-y^2}$, is shifted from $-$0.14
(LDA) to $-$0.03 (QSGW) while E$_{3z^2-r^2}$ is from $-$0.98 (LDA) to
$-$1.68 (QSGW) (indicated by the arrows in Fig.~\ref{Figure 1}; see also
Table~\ref{tab:1}). It is a factor of two difference in
$\Delta$E$_{e_g}$; 0.84 eV in LDA and 1.66 in QSGW. 
The correct estimation of this quantity is important especially in the
two-band theory for T$_c$ \cite{Sakakibara_PRL_2010,Sakakibara_PRB_2012_1,Sakakibara_PRB_2012_2,
  Sakakibara_PRB_2014}.
The large value of $\Delta E_{e_g}$ is indeed consistent with the RIXS 
data as will be discussed further below.

The same features are also found in HgBa$_2$CuO$_4$,
as presented in 
Fig.~\ref{Figure 1}(c) and (d). The $d_{x^2-y^2}$ band width is reduced
by $\sim$0.75 eV in QSGW compared to LDA and its center position of
PDOS moves slightly upward by 0.15 eV. While the $d_{3z^2-r^2}$
dispersion in this material is already quite small due to the thicker
blocking layer, its band width in QSGW is further reduced.  
$\Delta$E$_{e_g}$ is 2.25 in QSGW, again noticeably larger than the
LDA value of 1.67 eV (see Table~\ref{tab:1}).

In QSGW, the $\Delta$E$_{e_g}$ of both La$_2$CuO$_4$ and HgBa$_2$CuO$_4$ is 
enhanced compared to LDA/GGA. How does this affect 
the theoretical estimation of T$_c$? 
First, it should be noted that these parameters  cannot be 
directly adopted as the inputs for the FLEX calculation.
This is because, in principle, the QSGW self energy should
be partially subtracted before we put it into any of many-body
calculations. While there is no well-defined prescription
yet for this kind of `double-counting' problem \cite{Nd-PRB}, 
the ``best'' $\Delta$E$_{e_g}$ that should be adopted in the FLEX evaluation of 
T$_c$ may be lying somewhere in between the QSGW and LDA/GGA values. 
This can provide better quantitative agreement with the experiment, especially 
in La$_2$CuO$_4$, for which the LDA/GGA value of $\Delta$E$_{e_g}$ is  
found to be too small to account for $T_c=40$K.

\begin{figure}[t]
\begin{center}
\includegraphics[width=12cm,angle=0]{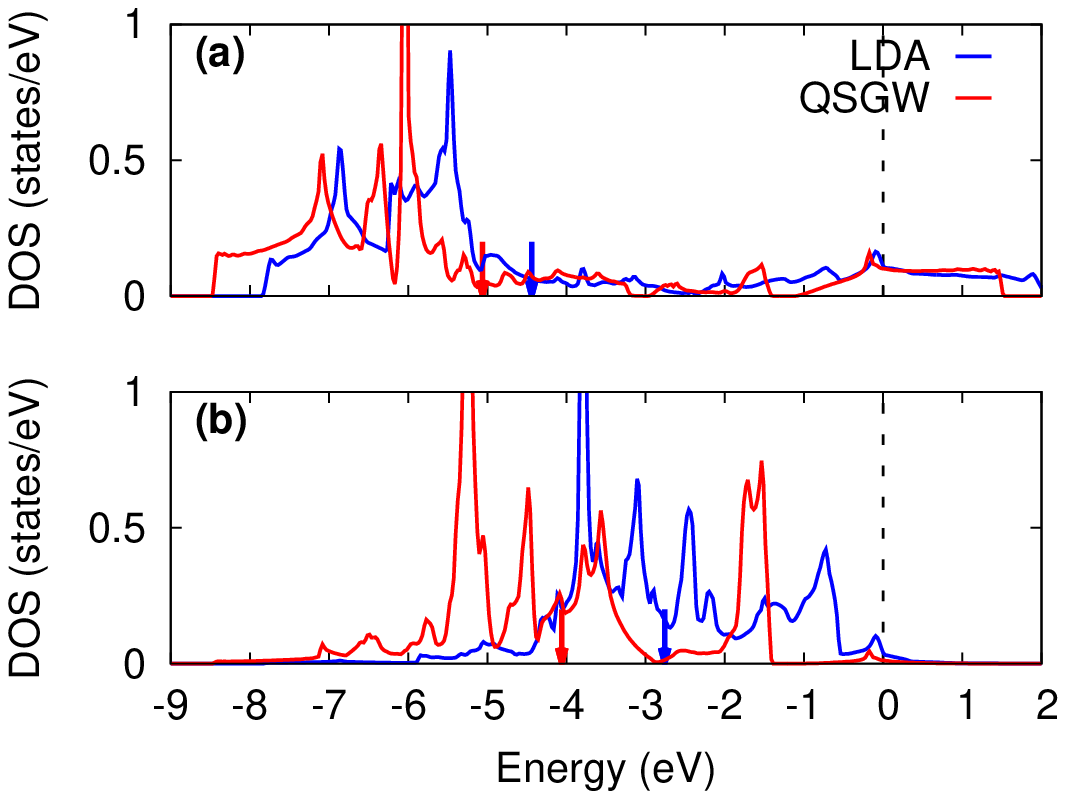}
\caption{% The calculated oxygen PDOS of La$_2$CuO$_4$
%  by LDA and QSGW. (a) In-plane oxygen $p_{x,y}^{\sigma}$ and (b)
%  out-of-plane oxygen $p_z^{\sigma}$ orbitals are plotted. The center
%  of mass position of each PDOS is marked by an
%  arrow. Fermi energy is set to be 0. 
\label{Figure 2}}
%PDOS-Op
\end{center}
\end{figure}

Oxygen states are also affected. Compared to LDA results, the O-$2p$
levels obtained by QSGW are significantly lowered in energy, as indicated in 
Fig.~\ref{Figure 2}. As summarized in Table~\ref{tab:1}, the center
position of in-plane oxygen PDOS is located at $-4.44$ ($-3.55$) eV in
LDA and at $-5.06$ ($-4.05$) eV in QSGW for La$_2$CuO$_4$
(HgBa$_2$CuO$_4$). The same feature is found for the apical O-$p_z$
PDOS. As a result, the energy difference, $\Delta$E$_p$=E$_{\rm
  apical}$--E$_{\rm inplane}$, is changed from 1.68 (1.00) in LDA to
0.55 ($-$0.86) in QSGW for the case of La$_2$CuO$_4$
(HgBa$_2$CuO$_4$), see Table~\ref{tab:1}. The correct estimation of
$\Delta$E$_p$ is also important for understanding T$_c$ since it is an
underlying quantity to determine $\Delta$E$_{e_g}$ ($\approx$
$\Delta$E $\approx$ $\Delta$E$_d$ + $\Delta$E$_{p}$ \cite{comment1})
in combination with other parameters.

Some other changes produced by QSGW are also noted. The $d_{3z^2-r^2}$
components in the bands below $-$1.5 eV in Fig.~\ref{Figure 1}(a) are
reduced in QSGW, and the free-electron-like bands at $\Gamma$ and $Z$
points above the Fermi energy are shifted upward. As the position of
the $t_{2g}$ complex is lowered (red color), the $d_{x^2-y^2}$ band has
almost no mixture with other bands below the Fermi
energy. Higher-lying La-$4f$ bands (not shown) move further upward as has been previously
noted in the nickelate systems \cite{Han-Kino-Kotani-PRB2014}.

\subsection*{Fermi surface}
The shape of the Fermi surface is important for understanding cuprate
superconductivity.  For example, its nesting is crucial for the spin
fluctuation pairing.  Also, a notable correlation between the
experimentally observed T$_c$ at the optimal doping (T$_c^{\rm max}$)
and the Fermi surface warping has been identified by Pavarini {\it et
  al.} \cite{Pavarini}. Here we discuss the Fermi
surface calculated by QSGW in comparison to the LDA result and experiment.

\begin{figure}[t]
\begin{center}
    \includegraphics[scale=0.6]{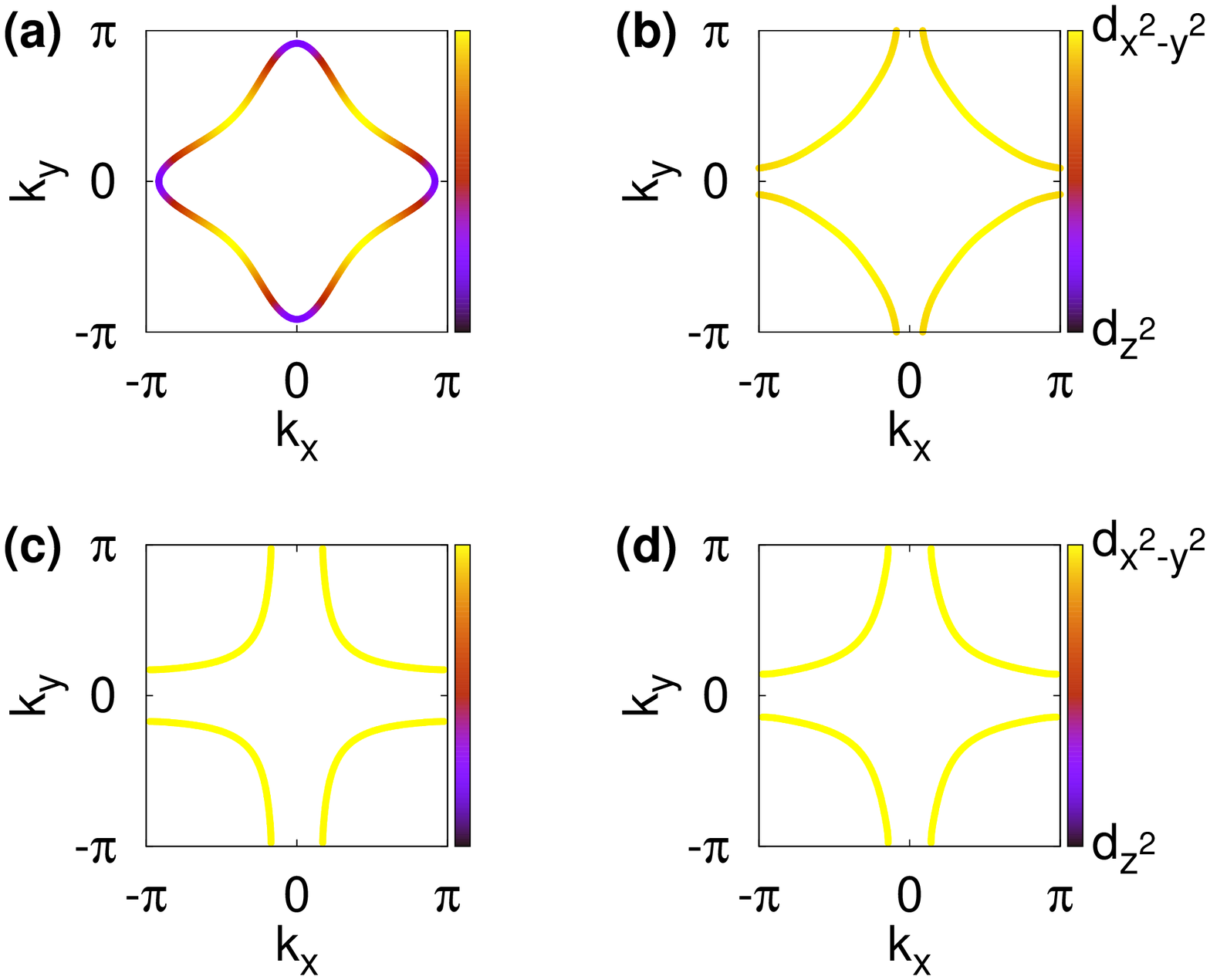}
    \caption{% The orbital-resolved
%        Fermi surfaces ($k_z$=0 plane) at the reduced $e_g$ band filling by 
%        0.15 (hole doping): (a) La$_2$CuO$_4$
%        (LDA), (b) La$_2$CuO$_4$ (QSGW), (c) HgBa$_2$CuO$_4$ (LDA),
%        and (d) HgBa$_2$CuO$_4$ (QSGW). The color represents the
%        amount of d$_{x^2-y^2}$ and d$_{3z^2-r^2}$ character.
}
 \label{Figure 3}
 %fig:Comparing_2D-Fermisurfaces
\end{center}
\end{figure}

The calculated Fermi surfaces are presented in
Fig.~\ref{Figure 3}; LDA ((a, c)) and QSGW
((b, d)). The hole doping is simulated by the rigid band shift method
so that the electron occupation in $e_g$ orbitals is reduced by
0.15$e$ per unit cell.  Notable features are found in the QSGW Fermi
surface for La$_2$CuO$_4$.  Contrary to the LDA result of
Fig.~\ref{Figure 3}(a),
Fig.~\ref{Figure 3}(b) has the pocket centered
at ($\pi$, $\pi$) point as in HgBa$_2$CuO$_4$ Fermi surface (see
Fig.~\ref{Figure 3}(c) and (d)).  This feature
is in good agreement with 
ARPES data~\cite{Ino} which also reports the pocket
centered at ($\pi$, $\pi$) point.  Further, the $d_{3z^2-r^2}$-orbital
character (dark purple) is significantly reduced and the $d_{x^2-y^2}$
character (bright yellow) is dominant in the QSGW result, which is
distinctive from the LDA  in which the significant amount of
$d_{3z^2-r^2}$ components are observed near ($\pi$, 0) and (0, $\pi$).

In the case of HgBa$_2$CuO$_4$, the difference between LDA and QSGW is
less pronounced, see Fig.~\ref{Figure 3}(c) and
(d).  While the QSGW Fermi surface is slightly more rounded, the
overall shape is not much different.  Since the $d_{x^2-y^2}$ orbital
character is dominant and $d_{3z^2-r^2}$ band is well separated from Fermi level
already in LDA due to the thicker blocking layers enhancing
two-dimensional feature, the LDA result is quite similar to QSGW.

\subsection*{Comparison with RIXS}

We now turn to the comparison with the RIXS data. Recently,
 Sala {\it et al.}~\cite{Sala} successfully extracted the
important model parameters for several different cuprate materials
based on RIXS spectra. In this subsection, we examine the material
dependent  parameters  by QSGW and
compare them to the experimental values.  With Ref.~\onlinecite{Sala}
as our main reference, we include two more compounds, namely,
Sr$_2$CuO$_2$Cl$_2$ and CaCuO$_2$ \cite{comment2}.

\begin{figure*}[ht]
\begin{center}
    \includegraphics[scale=0.4]{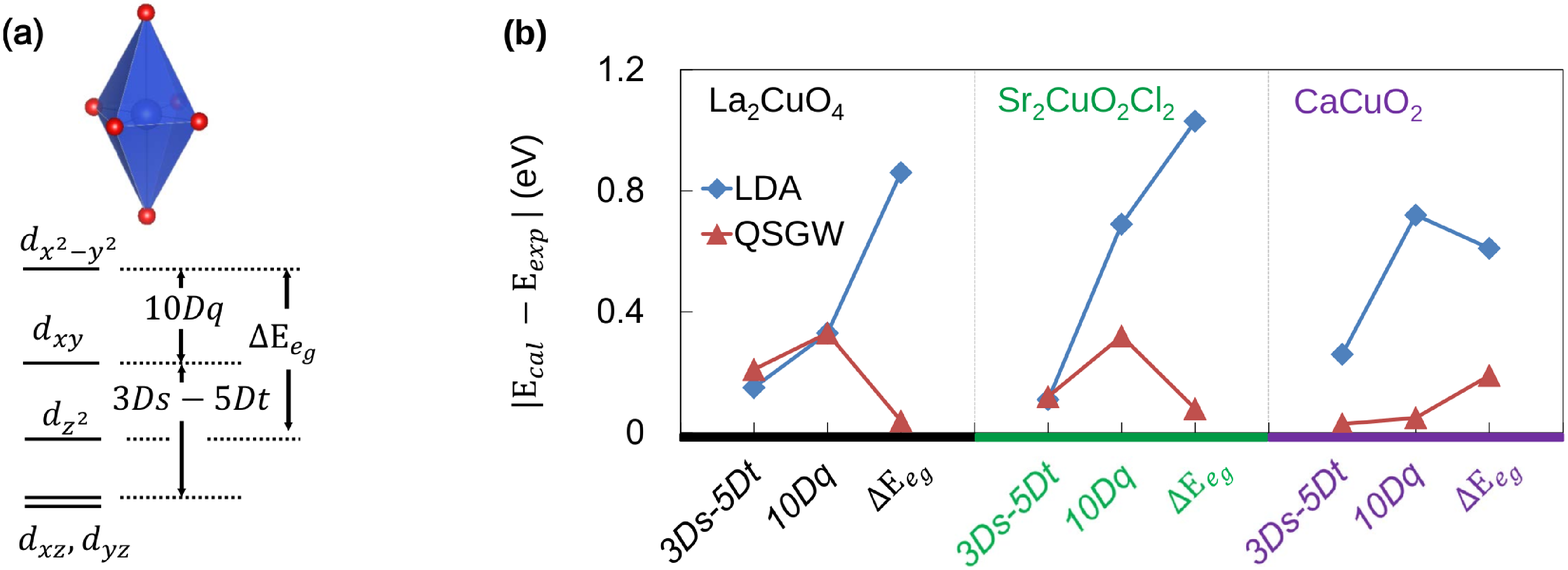}
    \caption{% (a) The three model parameters for the
%      comparison to Ref.~\onlinecite{Sala}. Note that ${\Delta}E_{e_g}$ is
%      denoted by $4Ds+5Dt$ in Ref.~\onlinecite{Sala}.  (b) The difference
%      of the calculated model parameters ($E_{\rm cal}$) from the experiments 
%      ($E_{\rm exp}$).
}
 \label{Figure 4}
 % fig:Model_Parameters
\end{center}
\end{figure*}

\begin{figure}[h]
\includegraphics[width=12cm,angle=0]{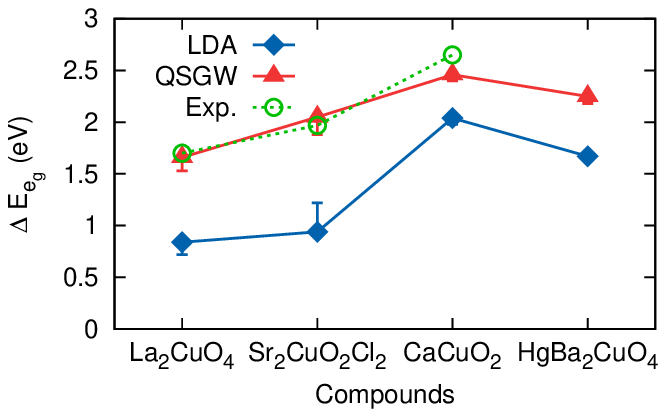} 
\caption{% The values of $\Delta$E$_{e_g}$ (or $4Ds +
%  5Dt$ in the notation of Ref.\onlinecite{Sala}) estimated by LDA (blue
%  squares), QSGW (red triangles), and RIXS data (green circles).
\label{Figure 5}}
% Eeg
\end{figure}

Our results are summarized in Table~\ref{tab:1},
Fig.~\ref{Figure 4}, and Fig.~\ref{Figure 5}.  The values of
10$Dq$, defined as the difference between two energy levels of
$d_{x^2-y^2}$ and $d_{xy}$ (see Fig.~\ref{Figure 4}(a)),
are larger in the QSGW calculation by $\sim$62--125\% than the LDA
values. While LDA underestimates $10Dq$ compared to the experiment,
QSGW slightly overestimates, which is related to the tendency that
Cu-$t_{2g}$ bands are pushed down relative to $e_g$, as was also observed
in the previous QSGW calculations for other transition-metal oxides
\cite{QSGW-PRL2006,QSGW-PRB2007,Han-Kino-Kotani-PRB2014}. It is
important to note that overall the QSGW result is in better agreement with
experiment, as clearly seen in Fig.~\ref{Figure 4}(b).

As noted in the above, according to the two-band theory by Sakakibara {\it et al.}
\cite{Sakakibara_PRL_2010,Sakakibara_PRB_2012_1}, 
the important parameter that governs T$_c$  is $\Delta$E$_{e_g}$ 
(or $4Ds$+$5Dt$ in Ref.~\onlinecite{Sala}).  
Fig.~\ref{Figure 4}(b) and
Fig.~\ref{Figure 5} clearly show that the calculated values of
$\Delta$E$_{e_g}$ by QSGW are in excellent agreement with those from
RIXS spectra; the difference is 2--8 \%. The LDA values are noticeably
smaller than the experiments although the difference gets reduced in the
higher T$_c$ materials, CaCuO$_2$ and HgBa$_2$CuO$_4$ (see Fig.~\ref{Figure 5}). 
This can be taken as a strong support for the two-band theory in the sense that
the LDA value of $\Delta$E$_{e_g}$ as an input for FLEX provides qualitative
information of material dependence, while the $\Delta$E$_{e_g}$ by QSGW
already contains the correlation effect beyond LDA, being consistent with RIXS.

Another parameter deduced from RIXS in Ref.~\onlinecite{Sala} is
$3Ds$--$5Dt$, the energy level difference between $d_{xy}$ and
$d_{yz,zx}$. In this case, the LDA results are not much different from
QSGW and experiment (see Fig.~\ref{Figure 4}(b)).

\subsection*{The effect of epitaxial strain}
An interesting aspect found in the T$_c$ trend of the cuprates is its
significant enhancement in the thin film form.  Locquet {\it et al.}
reported \cite{Locquet} that T$_c$ can be controlled by epitaxial strain
by about factor of two \cite{Bozovic}. The underdoped La$_2$CuO$_4$
with its bulk T$_c$ of 25K exhibits a higher and lower T$_c$ of
$\sim$49 K and 10 K when it is grown on SrLaAlO$_4$ (SLAO) and
SrTiO$_3$ (STO) substrates, respectively \cite{Locquet}.  It is
therefore important to check whether the two-band theory  is also
consistent with this observation.

In order to simulate the tensile and compressive strain produced by STO and
SLAO, we first optimized the $c$ lattice parameter with two different
in-plane lattice constants, $a^{\rm STO}$=3.905 and $a^{\rm
  SLAO}$=3.755~\AA, for La$_2$CuO$_4$, which originally has
$a_0$=3.782~\AA~ and $c_{0}$=13.25~\AA.  As expected, the optimized
out-of-plane parameters get smaller and larger under the tensile and
compressive strain, respectively; $c_{\rm STO}^{\rm
  opt}$=12.96 and $c_{\rm SLAO}^{\rm opt}$ = 13.36{\AA}. As a result,
the ratio between the out-of-plane and in-plane Cu--O distance,
$r=d_{\rm apical}/d_{\rm inplane}$, is found to be 1.32, 1.28, and
1.24, for $a^{\rm SLAO}$, $a_0$ and $a^{\rm STO}$,
respectively.

\begin{figure}[htbp]
\begin{center}
    \includegraphics[scale=2]{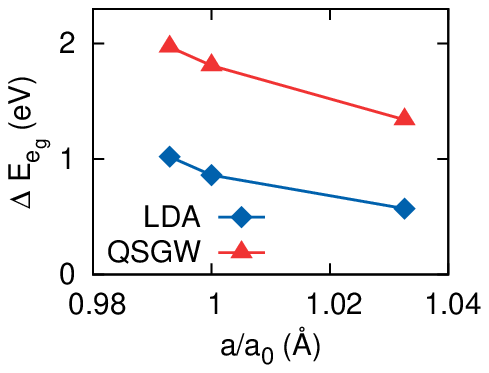}
    \caption{% The calculated ${\Delta}E_{e_g}$ as a
%      function of epitaxial strain. The $a_0$=3.782~\AA~ is the
%      experimental value for bulk La$_2$CuO$_4$. The compressive and
%      tensile strain are simulated with $a=3.755$ and 3.905~\AA~
%      considering the substrate of SrLaAlO$_4$ and SrTiO$_3$,
%      respectively \cite{Locquet}.
}
 \label{Figure 6}
 % fig:Thin_Film
\end{center}
\end{figure}

The calculated values of $\Delta$E$_{e_g}$ are plotted in
Fig.~\ref{Figure 6}.  Both LDA and QSGW predict that
$\Delta$E$_{e_g}$ gets enhanced and reduced under compressive and
tensile strain, respectively, which is consistent with the experimental
observation \cite{Locquet}. The reduction of $\Delta$E$_{e_g}$ at
a=$a_{\rm STO}$ is about 0.16 eV in both LDA and QSGW, and the
enhancement at a=$a_{\rm SLAO}$ is 0.29 (LDA) and 0.47 eV (QSGW).

\section*{Summary and Conclusion}

Using the QSGW method, we re-examined the electronic structure of copper
oxide high temperature superconducting materials.  Several important
features were found to have been captured by the $GW$ procedure, such as effective
mass enhancement. The shape and orbital character of the Fermi surface were
also notably changed,  especially for the case of
La$_2$CuO$_4$, and they are in good agreement with 
the ARPES data \cite{Ino}. 
Important model parameters including  the key quantity for the two-band theory of T$_c$, $\Delta$E$_{e_g}$,
were examined, and the QSGW results were in excellent agreement with RIXS data.

%The LDA values, used as inputs of the many-body calculation of 
%superconductivity in Sakakibara's theory, are found to be smaller compared 
%to RIXS. However, this should not be taken as an evidence against the theory, 
%since the RIXS experiment should be compared with the 
%output of the many-body calculation.

The present study shows that the first-principles 
band calculation can quantitatively reproduce the 
experimental observation by taking into account the correlation effects beyond 
LDA. We emphasize that it is not inconsistent with the previous study by Sakakibara {\it et al.}
which takes the LDA result as an input for the many-body calculation of superconductivity.
While the QSGW result cannot be used as a direct input for the FLEX-type 
calculation because of the partial double-counting of the many-body 
correlation, the ``best'' non-interacting Hamiltonian, 
that can serve as an input, may lie somewhere in between the 
LDA and QSGW. Obtaining a well-defined non-interacting Hamiltonian is, therefore,
an important future direction for the first-principles-based description of
high-temperature superconductivity, and it may quantitatively resolve the 
problem of  low T$_c$  in La$_2$CuO$_4$ produced by the  LDA input \cite{MiyaharaArita}.

\section*{Methods}

\subsection*{Quasiparticle self-consistent $GW$}
The QSGW \cite{QSGW-PRL2004,QSGW-PRL2006,QSGW-PRB2007} 
calculates $H_0$ (non-interacting Hamiltonian describing
quasiparticles or band structures) and $W$ (dynamically-screened
Coulomb interactions between the quasiparticles within the random
phase approximation) in a self-consistent manner.  
While the `one-shot' $GW$ is a perturbative calculation starting from a
given $H_0$ (usually from 
LDA/GGA), QSGW is a self-consistent perturbation method that can
determine the one-body Hamiltonian within itself.
The $GW$ approximation gives the one-particle
effective Hamiltonian whose energy dependence comes from the
self-energy term $\Sigma(\omega)$ (here we omit index of space and
spin for simplicity), and in QSGW, the static one-particle potential
$\vxc$ is generated as
\begin{eqnarray}
\vxc = \frac{1}{2}\sum_{ij} |\psi_i\rangle \left\{ {{\rm
    Re}[\Sigma(\ei)]_{ij}+{\rm Re}[\Sigma(\ej)]_{ij}} \right\}
\langle\psi_j|,
\label{eq:vxc}
\end{eqnarray}
where $\ei$ and $|\psi_i\rangle$ refer to the eigenvalues and
eigenfunctions of $H_0$, respectively, and ${\rm
  Re}[\Sigma(\varepsilon)]$ is the Hermitian part of the self-energy
\cite{QSGW-PRL2004,QSGW-PRL2006,QSGW-PRB2007}.  With this $\vxc$, one
can define a new static one-body Hamiltonian $H_0$, and continue to
apply $GW$ approximation until converged.  In principle, the final
result of QSGW does not depend on the initial conditions.  Previous
QSGW studies, ranging from semiconductors
\cite{QSGW-PRL2006,QSGW-PRB2007} to the various $3d$ transition metal
oxides \cite{QSGW-PRL2006,QSGW-PRB2007,QSGW-SpinWave-JPCM2008} and
$4f$-electron systems \cite{QSGW-4f-PRB2007}, have demonstrated its
capability in the description of weakly and strongly correlated
electron materials.

\subsection*{Computation details}
We used our new implementation of QSGW \cite{PMTQSGW-kotani} by
adopting the `augmented plane wave (APW) + muffin-tin orbital (MTO)',
designated by `PMT' \cite{kotani_pmt_2010,kotani_pmt_2013}, for the
one-body solver.  The accuracy of this full potential PMT method is
proven to be satisfactory in the supercell calculations of
homo-nuclear dimers from H$_2$ through Kr$_2$ with the significantly
low APW energy cutoff of $\sim$ 4 Ry, by including localized MTOs
\cite{kotani_pmt_2013}. A key feature of this scheme for QSGW is that
the expansion of $\vxc$ can be made with MTOs, not APWs, which enables
us to make the real space representation of $\vxc$ at any ${\bf k}$
point.

We performed the calculations with the experimental crystal structures
\cite{Miller,Jorgensen,Wagner,Qin}, and used 10$\times$10$\times$10,
12$\times$12$\times$12, 12$\times$12$\times$8, and
14$\times$14$\times$14 ${\bf k}$ points for LDA calculations of
Sr$_2$CuO$_2$Cl$_2$, La$_2$CuO$_4$, HgBa$_2$CuO$_4$, and CaCuO$_2$,
respectively.  As for QSGW calculations, in order to reduce the
computation cost, the number of ${\bf k}$ points were reduced to be
5$\times$5$\times$5, 7$\times$7$\times$7, 8$\times$8$\times$4, and
8$\times$8$\times$8 for the first Brillouin zone of
Sr$_2$CuO$_2$Cl$_2$, La$_2$CuO$_4$, HgBa$_2$CuO$_4$, and CaCuO$_2$,
respectively. The MTO radii used in our calculations were as follows: (i)
1.58, 1.04, 0.89, and 1.38 {\AA} for Sr, Cu, O, and Cl in
Sr$_2$CuO$_2$Cl$_2$, (ii) 1.43, 0.97, and 0.86 {\AA} for La, Cu, and O
in La$_2$CuO$_4$, (iii) 1.10, 1.59, 1.05, and 0.83 {\AA} for Hg, Ba,
Cu, and O in HgBa$_2$CuO$_4$, and (iv) 1.54, 1.01, and 0.86 for Ca,
Cu, and O in CaCuO$_2$.

Many of the key parameters in this study are defined in terms of the energy
levels of each orbital, such as E$_{x^2-y^2}$ and E$_{3z^2-r^2}$.  To
quantify them we simply take the center of mass position of PDOS:
\begin{align}
\begin{split}
E_\alpha = \frac{\int_{E_{\rm min}}^{E_{\rm max}}{Eg_{\alpha}(E)dE}}{\int_{E_{\rm min}}^{E_{\rm max}}{g_{\alpha}(E)dE}},\\
\end{split}
\end{align}
where $g_{\alpha}(E)$ is PDOS for a given orbital $\alpha$. An
ambiguity is inevitably introduced in determining E$_{\rm min,max}$,
and we set the range to cover the whole antibonding band complex for
Cu-$e_g$ states. (E$_{\rm min}^{e_g}$, E$_{\rm max}^{e_g}$) for
La$_2$CuO$_4$ is ($-1.95$ eV, 2.05 eV) in LDA and ($-2.20$, 1.55) in
QSGW.  For HgBa$_2$CuO$_4$, the band dispersion changes and the values
of E$_{\rm min}^{e_g}$ and E$_{\rm max}^{e_g}$ are redefined
accordingly: (E$_{\rm min}^{e_g}$, E$_{\rm max}^{e_g}$)=($-2.40$,
2.50) in LDA and ($-2.55$, 1.65) in QSGW. Importantly, none of the
reasonably defined energy ranges change our conclusion
\cite{comment0}, and the values are well compared with those reported
in the previous study using a maximally localized Wannier function
\cite{Sakakibara_PRL_2010,Sakakibara_PRB_2012_1}.

\section*{Acknowledgments}
We thank Ryotaro Arita for helpful comment and Prof. Hiroshi
Katayama-Yoshida for hosting the helpful discussion. S.W.J. and M.J.H.
were supported by Basic Science Research Program through the National
Research Foundation of Korea(NRF) funded by the Ministry of
Education(2014R1A1A2057202).  The computing resource is supported by
National Institute of Supercomputing and Networking / Korea Institute
of Science and Technology Information with supercomputing resources
including technical support (KSC-2014-C3-050) and by Computing System
for Research in Kyushu University.

\section*{Author contributions}
S.W.J., T.K., and H.K. performed the calculations. 
All authors contributed in analyzing the results and writing the paper.

\section*{Additional information}
The authors declare that they have no competing financial interests.

\clearpage

\section*{Tables}

\begin{table*}[h]
\caption{The calculated parameters by LDA and QSGW.  The $p_{x,y}$ is from the in-plane
  oxygen, and p$_z$ from the out-of-plane. The definitions
  of parameters can be found in
  Fig.~\ref{Figure 4}. The experimental
  values are taken from
  Ref.~\onlinecite{Sala}.}
\vspace{0.3cm}
\renewcommand{\arraystretch}{1.3}

\begin{tabular}{m{1.9cm}
>{\centering}m{1.6cm}>{\centering}m{1.6cm}>{\centering}m{1.6cm}>{\centering}m{1.6cm}
>{\centering}m{1.6cm}>{\centering}m{1.6cm}>{\centering}m{1.6cm}>{\centering\arraybackslash}m{1.6cm}}
\hline\hline
						& \multicolumn{2}{r}{E$_{d_{x^2-y^2}}$ $-$ E$_{\rm Fermi}$ (eV)} & \multicolumn{2}{c}{E$_{d_{z^2}}$ $-$ E$_{\rm Fermi}$ (eV)} & \multicolumn{2}{c}{E$_{O_{p_{x,y}}}$ $-$ E$_{\rm Fermi}$ (eV)} & \multicolumn{2}{c}{E$_{O_{pz}}$ $-$ E$_{\rm Fermi}$ (eV)}\\\cmidrule[0.1pt](l{0.75em}r{0.75em}){2-3}\cmidrule[0.1pt](l{0.75em}r{0.75em}){4-5}\cmidrule[0.1pt](l{.75em}r{.75em}){6-7}\cmidrule[0.1pt](l{.75em}r{.75em}){8-9}	
						& LDA & QSGW & LDA & QSGW & LDA & QSGW & LDA & QSGW\\\hline
	Sr$_2$CuO$_2$Cl$_2$	& $-$0.29 & 0.01 & $-$1.23 & $-$2.05 & $-$4.67 & $-$4.92 & $-$3.61 & $-$5.44\\	
	La$_2$CuO$_4$		& $-$0.14 & $-$0.03 & $-$0.98 & $-$1.68 & $-$4.44 & $-$5.06 & $-$2.76 & $-$4.06\\ 
	HgBa$_2$CuO$_4$		& $-$0.20 & $-$0.05 & $-$1.87 & $-$2.30 & $-$3.55 & $-$4.05 & $-$3.00 & $-$4.91\\
	CaCuO$_2$			& $-$0.30 & $-$0.17 & $-$2.34 & $-$2.63 & $-$3,67 & $-$4.05 & None & None \\
\end{tabular}
\vspace{0.5cm}
\begin{tabular}{m{1.9cm}
>{\centering}m{1.4cm}>{\centering}m{1.4cm}>{\centering}m{1.6cm}>{\centering}m{1.4cm}
>{\centering}m{1.4cm}>{\centering}m{1.6cm}>{\centering}m{1.4cm}>{\centering}m{1.4cm}>{\centering\arraybackslash}m{1.6cm}}
\hline
						& \multicolumn{3}{c}{ $\Delta$E$_{e_g}$ (4Ds + 5Dt) (eV)} & \multicolumn{3}{c}{3Ds $-$ 5Dt (eV)} & \multicolumn{3}{c}{10Dq (eV)}\\\cmidrule[0.1pt](l{.75em}r{.75em}){2-4}\cmidrule[0.1pt](l{.75em}r{0.75em}){5-7}\cmidrule[0.1pt](l{.75em}r{.75em}){8-10}
						& LDA & QSGW & Exp\vspace{-0.2cm}\par(Ref.~\onlinecite{Sala}) & LDA & QSGW & Exp\vspace{-0.2cm}\par(Ref.~\onlinecite{Sala}) & LDA & QSGW & Exp\vspace{-0.2cm}\par(Ref.~\onlinecite{Sala})\\
						\hline
	Sr$_2$CuO$_2$Cl$_2$	& 0.94 & 2.05 & 1.97 & 0.22 & 0.21 & 0.33 & 0.81 & 1.82 & 1.50 \\	
	La$_2$CuO$_4$		& 0.84 & 1.66 & 1.70 & 0.17 & 0.11 & 0.32 & 1.47 & 2.13 & 1.80 \\ 
	HgBa$_2$CuO$_4$		& 1.67 & 2.25 & None & 0.42 & 0.42 & None & 0.90 & 1.57 & None \\
	CaCuO$_2$			& 2.04 & 2.46 & 2.65 & 0.05 & 0.28 & 0.31 & 0.92 & 1.69 & 1.64 \\
	\hline\hline
\end{tabular}

\label{tab:1}
\end{table*}

\clearpage

\section*{Figure Legends}

FIG.~\ref{Figure 1}: The band dispersion and PDOS of La$_2$CuO$_4$ (a, b)
  and HgBa$_2$CuO$_4$ (c, d) calculated by (a, c) LDA and (b, d) QSGW.
The green and blue colors refer
  to the $d_{x^2-y^2}$ and $d_{3z^2-r^2}$ characters while the size of
  the colored dots represents their weight. The same color scheme was
  used for PDOS. Other bands than the two $e_g$ states are represented
  by red color. The center of mass position of PDOS is marked by an
  arrow. Fermi energy is set to be 0.\\

FIG.~\ref{Figure 2}: The calculated oxygen PDOS of La$_2$CuO$_4$
  by LDA and QSGW. (a) In-plane oxygen $p_{x,y}^{\sigma}$ and (b)
  out-of-plane oxygen $p_z^{\sigma}$ orbitals are plotted. The center
  of mass position of each PDOS is marked by an
  arrow. Fermi energy is set to be 0. \\

FIG.~\ref{Figure 3}: The orbital-resolved
        Fermi surfaces ($k_z$=0 plane) at the reduced $e_g$ band filling by 
        0.15 (hole doping): (a) La$_2$CuO$_4$
        (LDA), (b) La$_2$CuO$_4$ (QSGW), (c) HgBa$_2$CuO$_4$ (LDA),
        and (d) HgBa$_2$CuO$_4$ (QSGW). The color represents the
        amount of d$_{x^2-y^2}$ and d$_{3z^2-r^2}$ character.\\

FIG.~\ref{Figure 4}:  (a) The three model parameters for the
      comparison to Ref.~\onlinecite{Sala}. Note that ${\Delta}E_{e_g}$ is
      denoted by $4Ds+5Dt$ in Ref.~\onlinecite{Sala}.  (b) The difference
      of the calculated model parameters ($E_{\rm cal}$) from the experiments 
      ($E_{\rm exp}$). \\
      
FIG.~\ref{Figure 5}: The values of $\Delta$E$_{e_g}$ (or $4Ds +
  5Dt$ in the notation of Ref.\onlinecite{Sala}) estimated by LDA (blue
  squares), QSGW (red triangles), and RIXS data (green circles).\\
     
FIG.~\ref{Figure 6}:  The calculated ${\Delta}E_{e_g}$ as a
      function of epitaxial strain. The $a_0$=3.782~\AA~ is the
      experimental value for bulk La$_2$CuO$_4$. The compressive and
      tensile strain are simulated with $a=3.755$ and 3.905~\AA~
      considering the substrate of SrLaAlO$_4$ and SrTiO$_3$,
      respectively \cite{Locquet}.

 \clearpage

\section*{Figures}

 \clearpage

 \clearpage

 \clearpage

 \clearpage

 \clearpage

\end{document}